\documentstyle[preprint,aps,eqsecnum]{revtex}

\title{Collective modes and the far-infrared absorption of the
two-dimensional electron gas in a periodic quantizing magnetic field}

\author{Andrei Manolescu$^*$}
\address{Institutul de Fizica \c{s}i Tehnologia Materialelor,\\
	 C.\ P.\ MG-7 Bucure\c{s}ti-M\u{a}gurele, Romania. }
\author{Vidar Gudmundsson}
\address{Science Institute, University of Iceland, Dunhaga 3,
	 IS-107 Reykjavik, Iceland.}

\tighten

\draft

\begin{document}

\maketitle

\begin{abstract}
We investigate the far-infrared (FIR) absorption of a two-dimensional
electron gas in a periodically modulated quantizing magnetic field. The
magnetic field varies along only one spatial direction and the external
time-dependent electric field is linearly polarized along that axis.
The mutual Coulomb interaction of the electrons is treated self-consistently
in the ground state and in the absorption calculation within the Hartree
approximation. The effects of the magnetic material on top of the
heterostructure as a grating coupler is included in the time-dependent
incident FIR electric field.   We show that similar to an electric modulation,
the absorption can be directly correlated to the underlying electronic
energy bands. In addition, the magnetic modulation leads to absorption
spectra with a richer structure due to the quite different static response
of the electron density to the modulation.
\end{abstract}

\vspace {7cm}

$^*$ Regular Associate of the International Centre for Theoretical
Physics, Trieste, Italy.


\section{Introduction}

There is a recent, increasing interest in fabricating and understanding
the behavior of semiconductor devices incorporating an electron gas, usually
planar, and a nonuniform, or periodic magnetic field (magnetic modulation)
of a nanometric length scale.
\cite{McCord90:2153,Carmona95:3009,Ye95:3013,Izawa95:706,Ye95:1441,Ye97:5444}
The basic methods consist in building metallic patterns on the top of the
device, of ferromagnetic or superconducting material, which in the presence
of a uniform external magnetic field give a position dependent contribution
to the total magnetic field in the plane of the two-dimensional electron
gas (2DEG).\cite{Gerhardts97:xxx}

Several theoretical papers have analyzed the commensurability oscillations of
the resistivity of such systems for weak magnetic fields,
i.e. in the classical regime,\cite{Xue92:5986,Peeters93:1466,Gerhardts96:11064}
by analogy with the Weiss
oscillations, occurring in the presence of a weak electric modulation.
For the quantum regime, of strong magnetic fields, magnetic nanostructures
like magnetic barriers and magnetic wells have been recently considered.
\cite{Peeters93:15166,Matulis94:1518}  For periodic magnetic fields in
two spatial directions another recent paper has discussed the energy
spectrum equivalent to Hofstadter's butterfly.\cite{Gerhardts96:9591}

So far only very few papers have been dedicated to many-body effects.  Wu and
Ulloa,\cite{Wu93:7182} have studied the electron density and the collective
excitations in a one-dimensional magnetic superlattice with a short period,
of the order of the magnetic length corresponding to the average magnetic field.
They have found a rapid transition from a two-dimensional to a one-dimensional
behavior of the 2DEG, with increasing amplitude of the magnetic modulation.
Such a calculation requires the inclusion of the density response of the
electron gas to the periodic magnetic field, self-consistently with the energy
spectrum. The energy spectrum consists of Landau bands, the degeneracy of the
former Landau levels being lifted by the nonuniform magnetic field.
Consequently, the electron density becomes nonuniform and reacts to the induced
electric field which tends to optimize the density fluctuations.  This
peculiar electrostatic response is strictly of a quantum mechanical origin;
in the classical limit, i.e. for low magnetic fields, there is no coupling
between a magnetic field and the charge density at equilibrium.  The quantum
effects occur for high magnetic fields, of few Tesla or
more, when the corresponding magnetic length is much smaller than the
modulation period.\cite{Gossmann97:xxx}

The time-dependent density response function and the collective excitations
of the 2DEG in the presence of a periodic electric or magnetic field have also
been discussed recently by Stewart and Zhang.
\cite{Stewart95:R17036,Stewart96:6019}
They calculate the response function in the random-phase approximation
finding subsingularities related to the transition energies
near the van Hove singularities (vHS) of the energy bands.  However, they
neglected the influence of the Coulombian electron-electron interaction
on the energy spectrum, and thus their results, even qualitatively correct,
are restricted to a very weak modulation amplitude, much smaller than the
cyclotron energy.

We have previously studied the FIR absorption by the magnetoplasmons
of the 2DEG subjected to a short period unidirectional electric modulation and
a uniform perpendicular magnetic field and discussed the effect of
the modulation on the absorption peaks.\cite{Manolescu97:xxx}  The peaks
specific to the homogeneous system acquire an internal structure which can be
related to the energy spectrum.  The periodic Landau bands have vHS
in the center and at the edges of the Brillouin zones.  For
a modulation period much longer than the magnetic length the Landau bands
are nearly parallel, with energy separation equal to the cyclotron energy.
In this case the absorption spectrum is nearly identical to that in the
homogeneous system.  For a shorter modulation period the bands cease to
be parallel, and the absorption peaks split into several sub-peaks following
the dominant excitation energy which occur near the vHS.
\cite{Stewart95:R17036,Stewart96:6019,Manolescu97:xxx} In the present paper
we want to discuss the absorption spectra in the presence of a magnetic
modulation.

We compare the magnetic modulation effects with the
situation of the homogeneous or electrically modulated 2DEG in the presence
of a uniform and perpendicular (external) magnetic field.  Therefore
we will always keep a uniform field component, $B_0$.  In the typical
experimental realizations the period of the spatial magnetic modulation
is of a few hundred nanometers.  We will thus consider a period
$a\gg\l_0$, where $l_0=(\hbar/eB_0)^{1/2}$.  Within this restrictions
we expect a dispersion of the excitation energies, determined by the
cyclotron energy corresponding to the local magnetic field.  In other
words, the Landau bands will be non-parallel even for a weak magnetic
modulation, and thus even a small nonuniform field component will have
important effects.  Therefore we expect FIR absorption spectra as rich
in structure as for a short period electric modulation.

For the numerical calculations the material constants are chosen for
GaAs: the effective mass $m_{eff}=0.067m$, the dielectric constant
$\kappa=12.4$, and we assume spin degeneracy.  The average electron
density is fixed to $\rho_0=1.446\,\,10^{11}$ cm$^{-2}$, such that
$\nu B_0=6$ Tesla, where $\nu$ is the filling factor.

\section{Modulation model and single-particle states}

We consider a 2DEG located in the plane ${\bf r}=(x,y)$, and an external
magnetic field whose component along the $z$ axis is independent of $y$,
but periodic in the $x$ direction, having the simple form
\begin{equation}
B(x)=B_0+B_1\cos Kx \, .
\label{mmod}
\end{equation}
We assume the 2DEG to be ideally thin, such that it does not feel any
in-plane component of the magnetic field.  We describe the system with the
Hamiltonian $H=H_0+V_H$, where $H_0=({\bf p} + e{\bf A})^2/2m_{eff}$ is
the noninteracting term, and $V_H$ is the Hartree potential, self-consistent
with the charge density.

In the Landau gauge, which is appropriate for our problem, the
vector potential is
\begin{equation}
{\bf A}=(0,B_0x+\frac{B_1}{K}\sin Kx)\,,
\end{equation}
and with this choice the wave functions of the electrons have the form
\begin{equation}
\psi_{nX_0}(x,y)=L^{-1/2}e^{-iX_0 y/l_0^2}\phi_{nX_0 }(x)\,,
\end{equation}
where $n=0,1,...$ and $X_0$ is the center coordinate. The $y$ coordinate is
thus isolated in a simple plane-wave, and $\phi_{nX_0 }$ are the
eigenfunctions of the reduced Hamiltonian
\begin{equation}
H_{0,x}=\hbar\omega_0\left[-\frac{l_0^2}{2}\frac{d^2}{dx^2}+
\frac{1}{2l_0^2}\left(x-X_0+\frac{s}{K}\sin Kx\right)^2\right] \,,
\label{reha}
\end{equation}
in which $s=B_1/B_0$ measures the modulation strength, and $\omega_0=
eB_0/m_{eff}$.

We will restrict ourselves to the case when $s<1$, that is the magnetic
field (\ref{mmod}) vanishes nowhere.  In this case we can expand the
wave functions $\phi_{nX_0 }$ in the basis of the Landau wave functions
(i.e. the eigenfunctions of $H_{0,x}$ for $s=0$) and we can neglect the
contribution of very high Landau levels.  Since we are interested in a
magnetic modulation with a period $a=2\pi/K$ much longer than
$l_0$, $Kl_0\ll 1$,
we expect the perturbed wave functions $\phi_{nX_0 }$ to preserve the
localized character of the Landau wave functions, and only their ``center of
weight'' eventually shifts from the center coordinate $X_0$ to a new position
$X_1$.\cite{Gossmann97:xxx}
The relation between $X_1$ and $X_0$ can be found by linearizing the
sine of Eq.(\ref{reha}), $\sin Kx\approx\sin KX_1+K(x-X_1)\cos KX_1$, and by
reconstructing a new parabolic potential centered on $X_1$ one gets
\begin{equation}
H_{0,x}\approx\hbar\omega_0\left[-\frac{l_0^2}{2}\frac{d^2}{dx^2}+
\frac{(1+s\cos KX_1)^2}{2l_0^2}(x-X_1)^2\right] \,,
\label{apreh}
\end{equation}
and
\begin{equation}
X_1=X_0-\frac{s}{K}\sin KX_1 \,.
\label{scc}
\end{equation}

The Hamiltonian (\ref{apreh}) is the same as
(\ref{reha}) with $s=0$, but with modified cyclotron frequency
$\tilde\omega_0=\omega_0(1+s\cos KX_1)$ and magnetic length
$\tilde l_0=l_0/\sqrt{1+s\cos KX_1}$.  The nonuniform magnetic field has
thus also a scaling effect on the width of the wave functions, periodically
in the center-coordinate space.  The Landau bands resulting from
Eq.(\ref{apreh}) are thus
\begin{equation}
E_{nX_0}=\left(n+\frac{1}{2}\right)\hbar\tilde\omega_0 \,,
\label{aplb}
\end{equation}
where one can directly see the spatial dispersion of the magnetic field,
Eq.(\ref{mmod}), and the Landau levels corresponding to the local cyclotron
energy $e(B_0+B_1\cos Kx)/m_{eff}$, as we anticipated in the Introduction.
The energy gap in the center and at the edges of the Brillouin zone, that
is in the region where the density of states has vHS, is
$\hbar\omega_{\pm}=\hbar e (B_0 \pm B_1)/m_{eff}$.

The electrostatic Hartree energy is determined by the charge distribution,
\begin{equation}
\rho({\bf r})=2\sum_{nX_0} {\cal F}(E_{nX_0})\mid\phi_{nX_0}(x)\mid^2=
\sum_{m\geq 0}\rho_m \cos mKx \,,
\label{pade}
\end{equation}
${\cal F}$ being the Fermi function, and the Coulomb potential,
\begin{equation}
V_H(x)=\frac{e^2}{\kappa}\int d{\bf r'} \frac{\rho({\bf r'})}
{\mid{\bf r}-{\bf r'}\mid}=\frac{2\pi}{K}\frac{e^2}{\kappa}
\sum_{m\geq 1} \frac{\rho_m}{m} \cos mKx \,.
\label{elpo}
\end{equation}
The system is assumed electrically neutral, and thus the average
density $\rho_0$ does not enter Eq.(\ref{elpo}), but only
determines the chemical potential contained in the Fermi function.
The effect of the Hartree potential is a strong reduction of the
density fluctuations imposed by the nonuniform magnetic field
and also a reduction of the width of the Landau bands
intersected by the Fermi level.\cite{Gossmann97:xxx}

In the numerical calculations we diagonalize iteratively the Hamiltonian
$H_{0,x}+V_H(x)$ self-consistently with  the particle density, by mixing
up to 10 Landau levels, and using up to 20 Fourier coefficients of the density.
In Fig.\ 1 and Fig.\ 2 we show two characteristic energy spectra and density
profiles for the self-consistent problem. We have fixed $B_1=0.2$ T.
The modulation period is also fixed, $a=500$ nm. We formally keep a
finite temperature in all the calculations, $T=1$ K.
In Fig.\ 1 $B_0=2$ T ($\nu=3$) such that the Fermi level is in the second
Landau band, which is very flat.  The dispersion of this energy band is
canceled due to the electrostatic response. In Fig.\ 2 we have the opposite
situation, in which, due to the low density of states at the Fermi level
the energy dispersion is broad.  But even if the Landau bands given by
Eq.\ (\ref{aplb}) may be strongly distorted due to the Hartree potential,
the excitation energies $E_{n+1,X_0}-E_{n,X_0}$ are still given to a good
approximation by $n\tilde\omega_0$. This is because the main action of the
periodic magnetic field component,
at least in the regime we are interested here, is
on the degeneracy of the original Landau levels.  Thus, for $Kl_0\ll 1$
we may speak about a local character (in space) of the Landau-level
degeneracy, as the mechanism of the static density response,
and thus for the induced electrostatic potential.

The self-consistent electron-density profile may be smooth, like in Fig.\ 1(b),
but also more complicated, with higher harmonics like in Fig.\ 2(b).
In the latter case, in the spatial regions corresponding to the intervals of the
Brillouin zone with an integer filling factor the density is constrained
to follow proportionally the external magnetic field, Eq.\ (\ref{mmod}),
while in the regions with non-integer filling factors, the filling factor
itself is adjusted by the electrostatic response, and the spatial variation
of the density and of the magnetic field may be opposite.
Another reason for the short-range oscillations of the static induced density is the
combined effect of the shift of the wave functions, described by
Eq.\ (\ref{scc}) and of the scaling of the wave function width from
$l_0$ to $\tilde l_0$. Both these effects are periodic but out of phase.
Consequently, short harmonics, with a characteristic length of a few
$l_0$ are present in the static density response.  Such oscillations are
most prominent for not fully occupied Landau bands, Fig.\ 2(b).
\cite{Gossmann97:xxx}

\section{Collective oscillations and absorption spectra}

We investigate the plasma oscillations and the energy absorption of the
system with the help of the dielectric matrix $\varepsilon_{GG'}(q,\omega)$,
$G$ and $G'$ being vectors in the reciprocal space, i.e. they belong to
the set $mK,\, m=0,\pm 1,\pm 2,...$\,.  We assume a weak time-dependent
linearly polarized
electric field incident on the system, of the form $ {\bf E}=(E_x,0)$,
$E_x$ having only one spatial Fourier component, of wave vector $q=(q_x,0)$,
and also only one temporal Fourier component, of frequency $\omega$.

In the most common experimental situation the electric field is modulated
by a system of grating couplers.\cite{Batke85:2367,Wulf94:17670}
In our case such a modulation, with $q_x\equiv q=K$, can be produced by the
magnetic or by the superconducting metallic strips themselves.

We calculate the dielectric matrix
in the random-phase approximation, which is consistent with the Hartree
approximation we adopted for the energy spectrum in the ground state, such that
\begin{equation}
\varepsilon_{GG'}(q,\omega)=\delta_{GG'}-
\frac{2\pi e^2}{\kappa | q+G |}\chi_{GG'}(q,\omega) \,,
\label{}
\end{equation}
where the dielectric susceptibility $\chi_{GG'}$ is given by the Lindhard
formula adapted to the periodic system,\cite{Wulf90:3113}
\begin{eqnarray}
      &&\chi_{GG'}(q,\omega)=\frac{1}{a\pi l^2}
      \sum_{nn'}\int_0^a dX_0\nonumber\\
      &&\times\frac{{\cal F}(E_{n,X_0})-{\cal F}(E_{n',X_0})}
      {E_{n,X_0}-E_{n',X_0}-\hbar\omega-i\eta}\nonumber\\
      &&\times\, {\cal J}_{nn';X_0}(q+G) {\cal J}^*_{nn';X_0}(q+G')\,.
\label{linfo}
\end{eqnarray}
Here $\eta\to 0^+$ is the adiabatic switching parameter and
${\cal J}_{nn';X_0}(q)=\langle \psi_{n',X_0}|e^{iqx}|
\psi_{n,X_0}\rangle$.

Due to the periodicity of the system
$\varepsilon_{GG'}(q,\omega)= \varepsilon_{G-K,G'-K}(q+K,\omega)$.
This property ensures the periodicity of the plasma frequencies,
i.e. the zeros of \, $\det\varepsilon_{GG'}$\, satisfy $\Omega(q)=\Omega(q+K)$,
leading to a multiplication of the branches corresponding to the homogeneous
system.

In the FIR spectroscopy however, only a few modes are observable, i.e.
those with a non-vanishing oscillator strength.  The absorbed power can be
calculated from the Joule law of heating, expressible
as\cite{Dahl90:5763}
\begin{equation}
      P(q,\omega)=-\frac{\omega}{4\pi}
      {\rm Im} \, \varepsilon^{-1}_{GG} (q_1;\omega)
      \,q\,| \phi_{ext}(q,\omega)|^2 \,,
\label{abspo}
\end{equation}
where $q=q_1+G$ with $q_1$ being the wave vector reduced to the first Brillouin
zone, $0\le q_1 \le K$, and $\phi_{ext}(q,\omega)$ denotes the electric
potential of the incident field.  For simplicity, we will  normalize
the external potential in Eq.(\ref{abspo}) such that
$\frac{1}{4\pi}q|\phi_{ext}(q,\omega)|^2=1$.  In order to evaluate the
absorbed power we will assume a certain dissipation in the system,
by using a finite adiabatic switching parameter in Eq.(\ref{abspo}),
$\eta=\hbar\omega_0/50$.

In Fig.\ 3 we compare the absorption for the homogeneous system with that for a
magnetically modulated system, for several wave vectors, with the same uniform
magnetic field $B_0=2$ T.  In the homogeneous system, for a low $q$, only
the lowest magnetoplasma mode is active, according to the well known
equation
\begin{equation}
      \omega^2=\omega_0^2+2\pi \rho_0 e^2 q/(\kappa m_{eff})\,.
\label{mdlq}
\end{equation}
With increasing $q$ the oscillator strength of the higher mode, around
$2\omega_0$, progressively increases, starting at about $q=0.4K$.  The high
modes at $n\omega_0$, $n=2,3,...$ are known as Bernstein
modes.\cite{Bernstein58:10,Gudmundsson95:17744}
In the modulated system we observe, in addition,
intermediate modes, with $\omega_0 < \omega <2\omega_0$, and some other
small peaks outside this interval.

In Fig.\ 4(a) we show the dispersion of the first two modes for the
non-periodic system versus the absolute wave vector $q$. In Fig.\ 4(b) we have
folded the curves up to $q=3.8K$ into the first Brillouin zone.
Due to the reflection symmetry of the
problem as imposed by the choice of the periodic magnetic field,
Eq.(\ref{mmod}), we have restricted Fig.\ 4(b) to half of the unit cell in the
wave-vector space, $0\le q_1 \le K/2$, the next half, corresponding to
$K/2\le q_1 \le K$, being the reflection of the first half. In the modulated
system one expects a deviation of the frequency curves from those of Fig.\ 4(b)
and a splitting at all the crossing points\cite{Wulf90:3113}.  However, the
peak structure of the traces in Fig.\ 3(b) can be completely understood with
Fig.\ 4(b).  For the modulated system several frequency branches (bands)
become active for any value of $q$.

As we already mentioned, in the periodic system only the frequency dispersion
is periodic with $q$, not the oscillator strength, which is the most
relevant quantity in the absorption spectroscopy.  For a low $q$ the modulated
system behaves indeed like the homogeneous one, but with increasing $q$, the
oscillator strength distributes to the neighouring branches.  For instance,
for $q=0.4K$ and $q=0.6K$, both corresponding to $q_1=0.4K$ in Fig.\ 4(b),
instead of one peak for $1.2\omega_0<\omega<1.4\omega_0$, belonging to the
lowest and respectively to the next frequency band shown in Fig.\ 4(b), we see
in Fig.\ 3(b) both frequencies active.  At $q=0.8K$ we still have a peak on 
the lowest band at $\omega\approx 1.1\omega_0$, 
at the same position as for $q=0.2K$, and also
several peaks on the upper branches.  The shoulder at about
$\omega=1.5\omega_0$ comes from the third band,
and it becomes the main peak at $q=1.2K$ ($q_1=0.2K$).  The frequency of the
main peak for $q=0.8K$ is no more resolved for $q=1.2K$, but the frequency of
the lowest branch is still observable.  We see further in each trace with
$q>0.8K$ two peaks, below and above $\omega=1.6\omega_0$ respectively, with
no dispersion in $q$.  They correspond to the splitting of the uppermost
branches, which in the homogeneous system are close to $\omega=1.6\omega_0$.
These branches are the most perturbed by the modulation: the lower is shifted
downwards and the upper is strongly pushed into the frequency gap, as shown by
the dashed lines in Fig.\ 4(b).  These dispersionless modes are the transition
modes from one to another Bernstein harmonic.

Similar dispersionless peaks are obtained for $\omega\approx 1.9\omega_0$,
$\omega\approx 2.0\omega_0$, and $\omega\approx 2.1\omega_0$, from the
branches with frequencies above $2\omega_0$.  The perturbation of these
branches due to the modulation is however stronger than on the lower branches,
below $2\omega_0$, because the excitation spectrum is now wider, between
$2\omega_-$ and $2\omega_+$, i.e.\ between $1.8\omega_0$ and $2.2\omega_0$.
Not all the observable peaks gain in intensity with increasing wave vector,
see for example the uppermost modes for $q=1.0K$, or $q=1.2K$.  In the latter
case one peak between two absorption maxima is completely missing, that is the
corresponding mode is inhibited (very low oscillator strength).

It is worth to mention that for the modulated system, in the gap region the
integral in Eq.(\ref{linfo}) is singular due to the poles of the integrand
between $2\omega_-$ and $2\omega_+$, such that a numerical calculation
of the gap modes as zeros of determinant of the dielectric matrix would become
very delicate.  Our analysis, with a finite imaginary part, is thus much more
efficient.

In Fig.\ 5 we show the evolution of the peak structure of the absorption
with increasing modulation amplitude, for $B_0=1.5$ T, that is for the
situation when the Fermi level is in a gap, and the Landau bands are wide.
Here we have fixed the wave vector $q=K$.   Like in Fig.\ 3, the modulation
activates the modes in the vicinity of the frequency  branches of the
unmodulated system.  All the frequencies are in fact changed by the modulation,
but the most significant change is again of those close to the frequency gap.
The modes originating at the gap edges are shifted into the gap.

The peaks which evolve directly from the peaks in the unmodulated
system can still be
distinguished by the structure of the time-dependent induced density, rather
than by the frequency position, as we want to demonstrate in Fig.\ 6 for
$B_1=0.2$ T.  In the upper panel we show the induced density for two
frequencies in the unmodulated system.  For $\omega=1.20\omega_0$ (not shown
in Fig.\ 5(a)) there is no absorption peak, and the particle density fluctuates
in time such that the time-dependent induced density is always symmetric.
That is no dipolar motion occurs within the modulation unit cell, $0<Kx<2\pi$.  On the contrary, for a peak frequency, like
$\omega=1.69\omega_0$, the induced density is antisymmetric, dipole active.
For the modulated system we first display, in the lower panel, a symmetric
induced density, for a frequency where no absorption takes place.
The symmetry is now only approximate, but the
absorption is insignificant at this frequency.  At $\omega=1.64\omega_0$ the
induced density reminds of the dipolar motion in the homogeneous system. It
is antisymmetric and the charge motion in the center of the unit cell is
spread over a large region. Due to the modulation some additional density
fluctuations are present at the edges of the cell.  For $\omega=1.52\omega_0$
the dominant motion is not far from the cell center, between $KX=1$ and $KX=4$,
while for $\omega=1.77\omega_0$ the density oscillations are confined at the
cell edges.  We ascribe this behavior to the specific energy spectrum.
The excitation frequencies split into $\omega_+$ and $\omega_-$
corresponding to the vHS of the energy bands, and thus to the flat regions in
the energy spectrum.  Since the modulation period is much larger than the
magnetic length, those magnetoplasma modes with frequencies slightly
below and slightly above the frequency in the homogeneous system, here
$\omega=1.52\omega_0$ and $\omega=1.77\omega_0$ respectively, are thus
determined by local oscillations of the electron density  around the center
and around the edges of the unit cell respectively.

The local character of the modes which are activated by the modulation is
better seen for the group around $2\omega_0$.  Even in the previous group
we already see some weak high harmonics in the induced density, for
$\omega=1.77\omega_0$. Such short range oscillations become very pronounced
for the higher frequencies.  Again the middle frequency, $\omega=2.11\omega_0$,
preserves the long range behavior of the peak mode in the unmodulated system.
For the lower frequency, close to $2\omega_-$, the center of the unit cell
is more active, this time with very pronounced short range harmonics, and
for the higher frequency, close to $2\omega_+$, the motion at the cell edges
becomes the most dipolar active.

\section{Comparison with an electric modulation}

For a comparison of the effects of a magnetic modulation with those
of a pure
electric modulation we show in Fig.\ 7 results obtained for $B_1=0$,
but with a periodic electrostatic potential $V\cos Kx$, again of period
500 nm, and of amplitude $V=20$ meV.  The uniform magnetic field is
$B_0=1.5$ T, and $B_0=2$ T, and the absorption spectra are represented
by the dashed and respectively the full lines of Fig.\ 7(a). The wave
vector of the incident field is again fixed to $q=K$.
In order to obtain significant deviations from the results in the
homogeneous system, comparable with those in the presence of the
periodic magnetic field, we had to increase the strength of the electric
modulation, $V/\hbar\omega_0$, almost two orders of magnitude above
the strength $s=B_1/B_0$ of the magnetic modulation.  The peak
structure of the absorption spectra in Fig.\ 7(a) is still less rich
than what we have shown previously for the magnetic modulation.

The reason is the much stronger static response of the density to the periodic
electrostatic potential than to the periodic magnetic field, compare
Fig.\ 7(b) with Figs.\ 1(b) and 2(b).  Consequently, the screening effects
in the presence of an electric modulation are much stronger than in the
presence of a magnetic modulation.  The Landau bands corresponding to
Fig.\ 7(a) have the same width as those in Fig.\ 1(a), for $B_0=2$ T,
and Fig.\ 1(b), for $B_0=1.5$ T.  The difference consists in weaker deviations
of the excitation energies from multiples of $\hbar\omega_0$ than
for the magnetic modulation, that is the Landau bands are now nearly parallel.
Consequently we see in the absorption spectra of Fig.\ 7(a) less modes,
i.e. those present in the homogeneous system for $q=K$, see Fig.\ 3(a) and
Fig.\ 5(a), plus only one or two additional modes from the neighboring branches
which are activated by the modulation, see Figs.\ 4(b) and 5(b).  The original
frequencies are significantly perturbed only for $B_0=1.5$ T, where we
observe the tendency of the marginal modes to penetrate the frequency gap,
respectively, the peak slightly above $\omega=1.7\omega_0$ and the shoulder
below $\omega=2\omega_0$.

The electron density induced by the incident field, shown in Fig.\ 7(c),
is long ranged for the lowest peaks, e.g. for $\omega=1.71\omega_0$,
resembling that in the unmodulated system.  Higher harmonics are now
present only for the small peak at $\omega=2\omega_0$, indicating the
onset of local charge oscillations.  Much weaker high harmonics are also
observable in the modes at the higher frequencies.  Since in the case of
the magnetic modulation the electron density has higher harmonics
already in the ground state, Fig.\ 2(b), short range dynamic fluctuations 
are favored in the presence of the incident field, and thus the
local modes, while for an electric modulation such modes are possible
only for a large modulation amplitude, or for a short period, i.e. for a
steep modulation potential.  To summarize, the influence of a long period
magnetic modulation on the FIR absorption in a 2DEG is qualitatively similar
to that of a short period electric modulation of a comparable strength.

\section{Conclusions}

We have analyzed the magnetoplasma modes perpendicular to a one-dimensional
magnetic modulation. We have discussed the influence of the Landau band
structure, the density profile, and self-consistent electrostatic screening
on the structure of the absorption spectra. We compare these effects
to the analog results obtained in the presence of an electric modulation.
In order to keep the interpretation as simple as possible, we have chosen
the simplest situation, when the incident electric field is oriented
perpendicular to the modulation, and is uniform along the modulation, that
is $q_y=0$.  This situation is usually encountered in experiments on grating
coupler devices.  In the magnetically modulated systems the modulation of the
incident, external electric field, with $q_x=K$, is possible by the metallic
strips already incorporated in the device for producing the periodic magnetic
field.  In order to explain the results we have also considered the more
general case with $q_x\neq K$.

Due to the weak and short range static density response of the 2DEG to a
periodic magnetic field, as compared to the response to a periodic
electrostatic field of the same period and strength, the screening effect
on the Landau band structure is much weaker in the former than in the latter
case.  Therefore the structure of the absorption spectra for the magnetic
modulation is much richer than for the electric modulation.  For simplicity,
we have limited our discussion to the situation when the modulation strength
is sufficiently weak, such that the absorption spectra can be related
to the collective modes occurring in the unmodulated system.

\acknowledgements
One of us (A.M.) wishes to thank International Centre for Theoretical Physics,
Trieste, Italy, for hospitality. This research was supported by the
Associateship Scheme of the ICTP, the Icelandic Natural Science
Foundation, and the University of Iceland Research Fund.

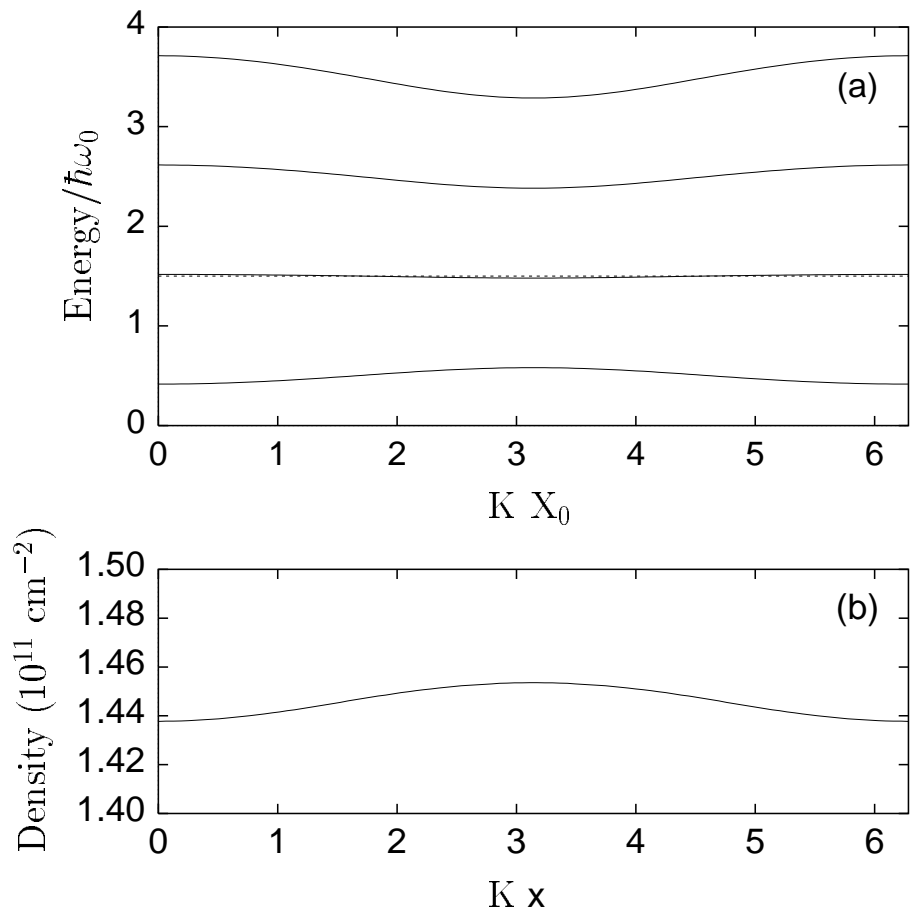
\begin{figure}
\caption{Energy bands (a) and electron density (b) for a magnetic modulation
with $B_0=2$ T and $B_1=0.2$ T.  The dashed line shows the Fermi level.}
\label{fig.1}
\end{figure}
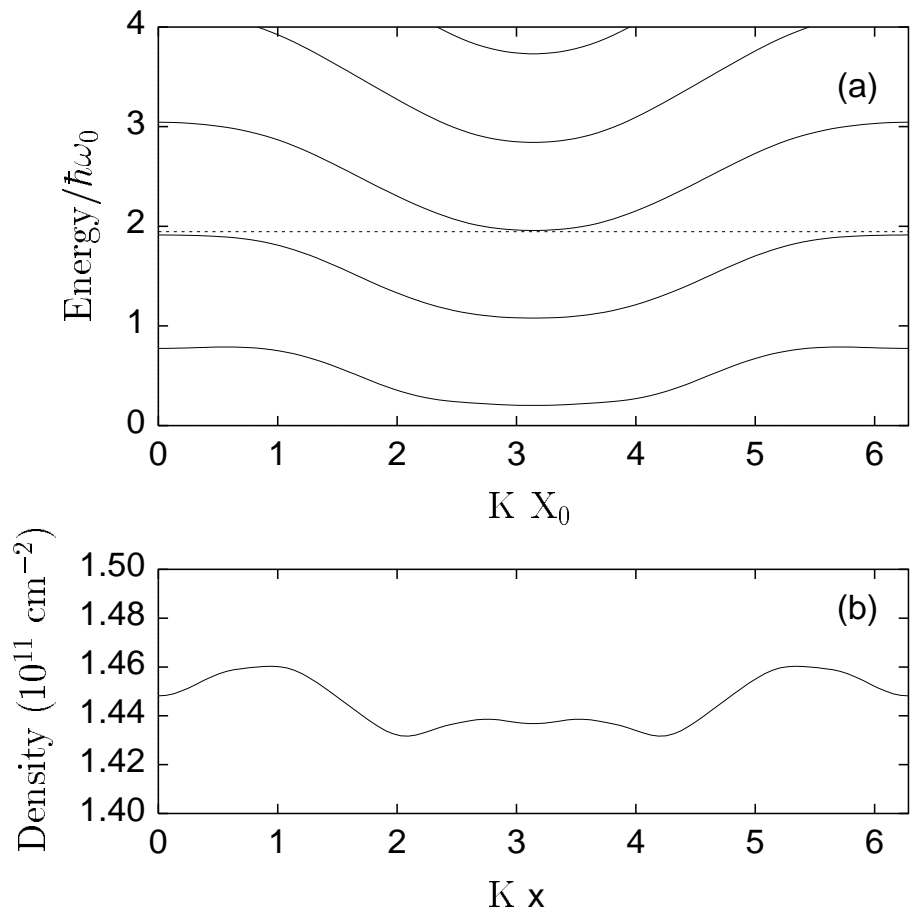
\begin{figure}
\caption{Energy bands (a) and electron density (b) for a magnetic modulation
with $B_0=1.5$ T and $B_1=0.2$ T.}
\label{fig.2}
\end{figure}
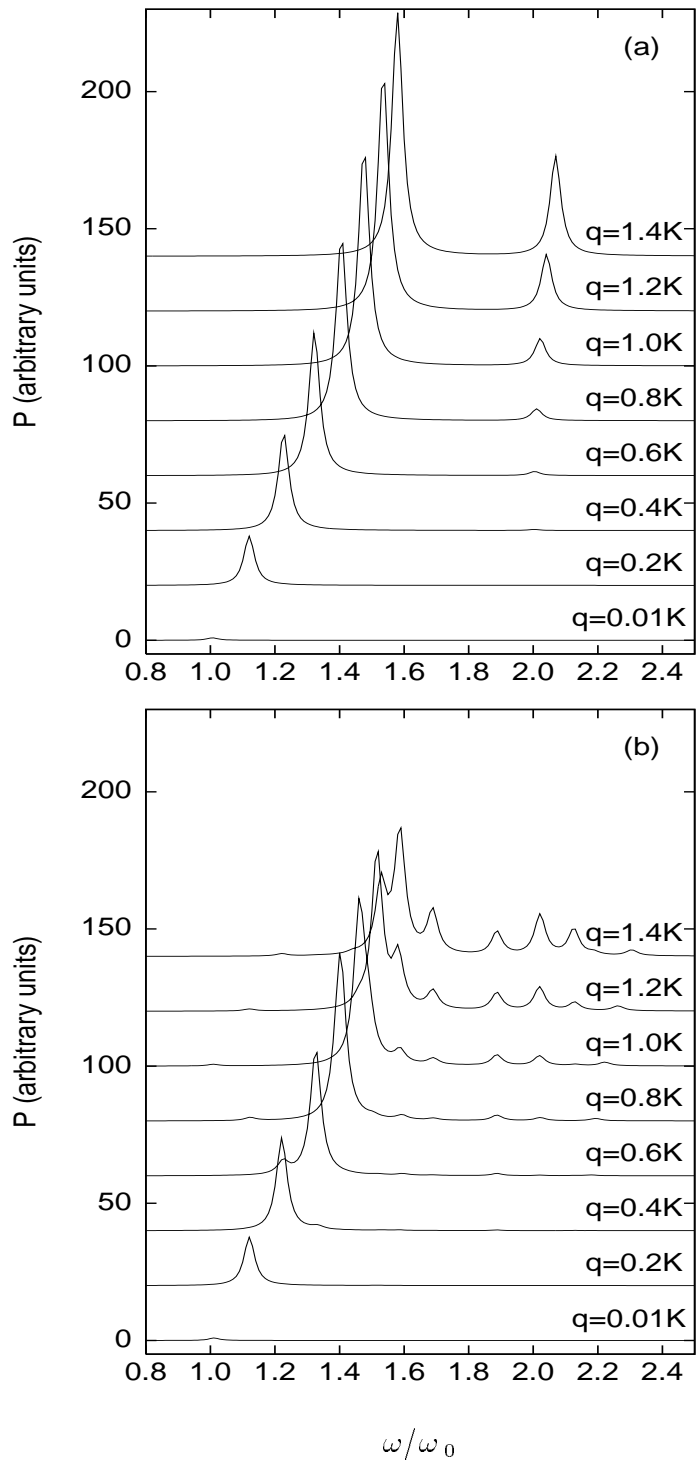
\begin{figure}
\caption{Absorption spectra for $B_0=2$ T in a homogeneous system (a)
and in a modulated magnetic field with $B_1=0.2$ T (b) for various
wave vectors of the incident electric field perpendicular to the
modulation.}
\label{fig.3}
\end{figure}
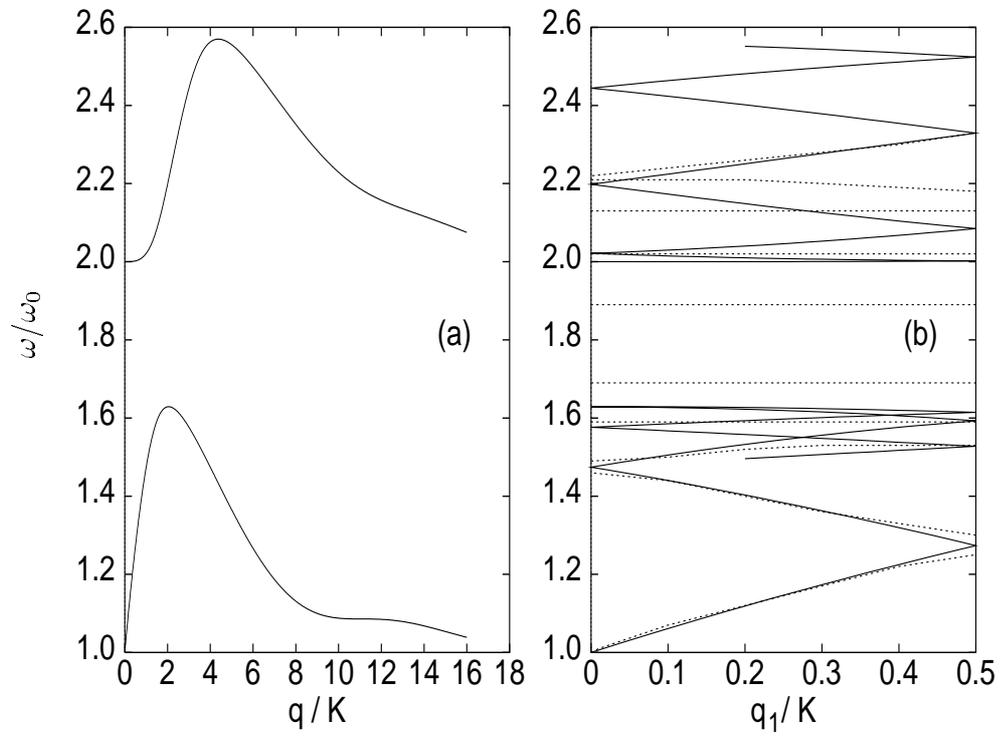
\begin{figure}
\caption{Magnetoplasma dispersion of the first two modes, in the homogeneous
system, for $B_0=2$ T (a) and the back-folded dispersions for $q\le3.8K$ (b)
where $q_1$ is the wave vector reduced to the interval $[0,K]$; due to the
reflection symmetry we have shown only the interval $[0,K/2]$.
The dashed line shows the dispersion of the observable absorption maxima in
the modulated system with $B_1=0.2$ T, for $q\le 1.5K$.}
\label{fig.4}
\end{figure}
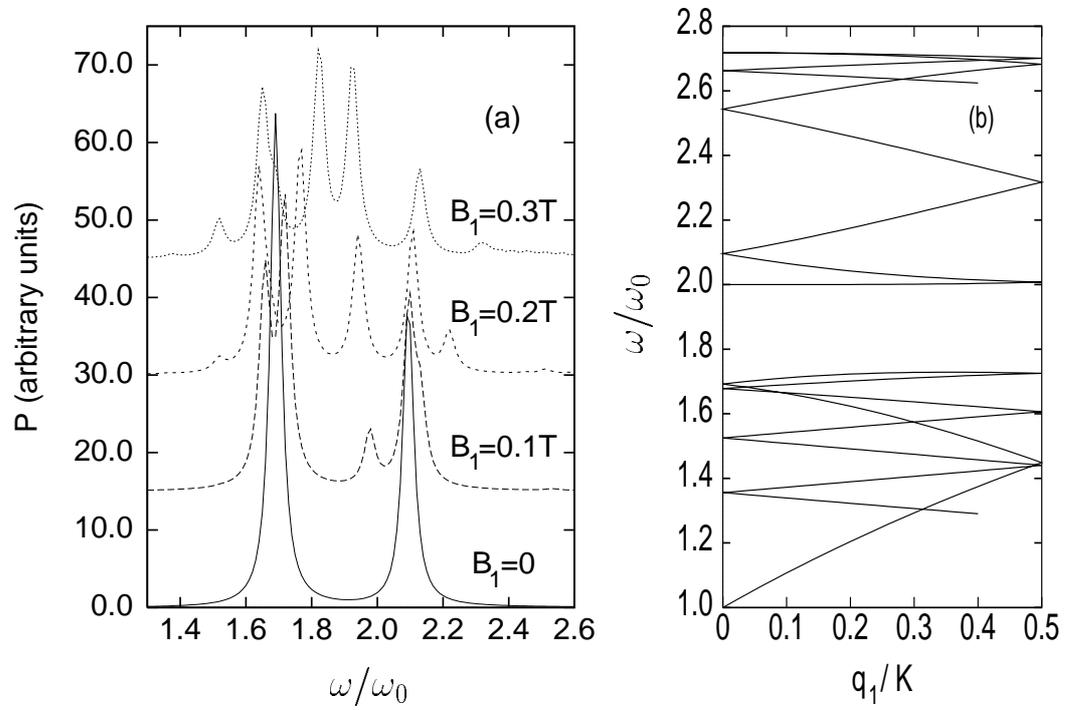
\begin{figure}
\caption{Various absorption spectra for a magnetic modulation with
$B_0=1.5$ T, and with an increasing $B_1$, with $q=K$ (a)
and the back-folded dispersions of the first two modes in the homogeneous
system, for $q\le4.4K$ (b). (Note that the dispersion depends on $B_0$,
being thus different from that of Fig.\ 4.)}
\label{fig.5}
\end{figure}
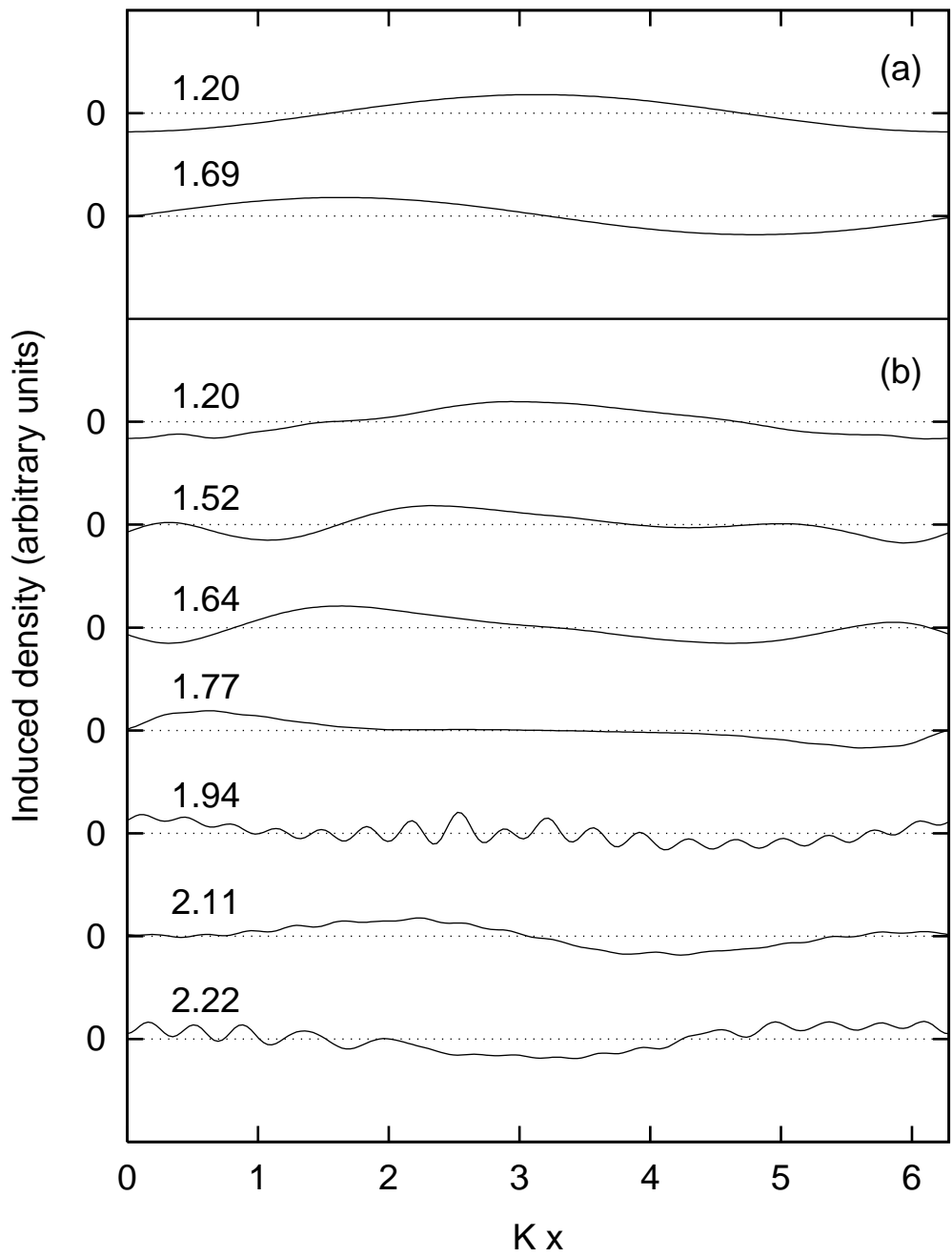
\begin{figure}
\caption{Induced electron density for $q=K$, for several frequencies shown
by the numbers inside the plots, in units of $\omega_0$,
for the homogeneous system (a) and for the
magnetic modulation with $B_0=1.5$ T and $B_1=0.2$ T (b).}
\label{fig.6}
\end{figure}
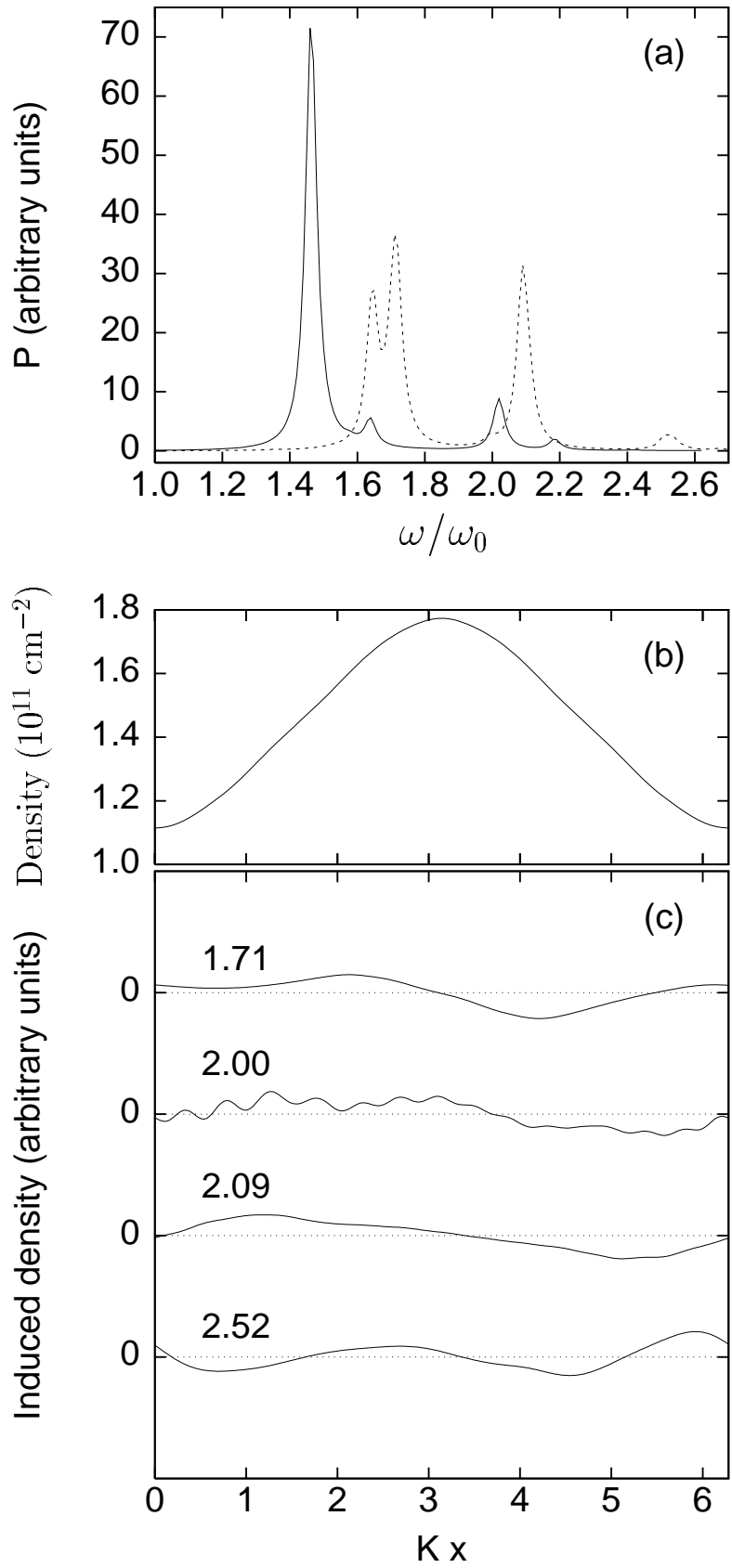
\begin{figure}
\caption{Results for a pure electric modulation of amplitude $V=20$ meV
($B_1=0$).  Absorption spectra for $q=K$, with $B_0=2$ T and $B_0=1.5$ T,
full and respectively dashed lines (a), electron density for $B_0=1.5$ T (b),
and a few induced densities for $B_0=1.5$ T, for the frequencies shown by
the numbers (in units of $\omega_0$) (c).}
\label{fig.7}
\end{figure}
\end{document}